# Nonstandard Convergence to Jamming in Random Sequential Adsorption: The Case of Patterned One-Dimensional Substrates

Arjun Verma  and  Vladimir Privman*

Department of Physics, Clarkson University, Potsdam, NY 13676, USA

## ABSTRACT

We study approach to the large-time jammed state of the deposited particles in the model of random sequential adsorption. The convergence laws are usually derived from the argument of Pomeau which includes the assumption of the dominance, at large enough times, of small landing regions into each of which only a single particle can be deposited without overlapping earlier deposited particles and which, after a certain time are no longer created by depositions in larger gaps. The second assumption has been that the size distribution of gaps open for particle-center landing in this large-time small-gaps regime is finite in the limit of zero gap size. We report numerical Monte Carlo studies of a recently introduced model of random sequential adsorption on patterned one-dimensional substrates that suggest that the second assumption must be generalized. We argue that a region exists in the parameter space of the studied model in which the gap-size distribution in the Pomeau large-time regime actually linearly vanishes at zero gap sizes. In another region, the distribution develops a threshold property, i.e., there are no small gaps below a certain gap size. We discuss the implications of these findings for new asymptotic power-law and exponential-modified-by-a-power-law convergences to jamming in irreversible one-dimensional deposition.

________________________________________________
*corresponding author; e-mail: privman@clarkson.edu

# 1. Introduction

Random sequential adsorption (RSA) is an important dynamical model [1-6] that describes irreversible deposition of particles or other objects on one-dimensional (1D) linear substrates, on two-dimensional (2D) surfaces, on scaffolds, etc. The objects are randomly transported to the substrate, but are attached only provided they do not overlap earlier-deposited objects. Once attached, the objects cannot move on the substrate or detach from it. Recently, there has been a renewed interest in the applications of RSA models [5,7-17] to pre-patterned 1D and 2D substrates. This has been in response to new experimental capabilities [18-43] to prepare micro- and nano-patterned substrates, including surfaces with well-defined preferential sites for specific particle attachment. In the studies of RSA one frequently focuses on the density and structure of the infinite-time jammed-state configuration. In applications, rapidly achieving dense coverage is usually preferred. Therefore, the asymptotic large-time laws describing approach to the jammed state coverage are of interest. Two standard prototype convergence laws have been found in extensively studied RSA models [1-6,9,44,45]. These include fast exponential vs. slow power-law approach to jamming as a function of time, $t$. The latter, power-law convergence can be modified by a power-of-a-logarithm factor [45] for some geometries of the depositing objects.

Recent technological advances have allowed tailoring the landing-site geometry for particle attachment, in addition to the earlier-studied effects of the particle shape and orientation, in order to control the RSA process. Presently, various growth and deposition processes have been experimentally realized on 1D lines [18-23] or nanotubes [24-26,43], etc., or 2D patterned surfaces [27-42]. Patterned surfaces find applications in electronics [24-28]; photovoltaics, optics, optoelectronics [29-32]; sensors, microarrays [33-38]; crystal growth and particle assembly [18-23,39].

The 1D RSA model provides a convenient test bench for studying the two convergence laws: exponential (fast) or power-law (slow). Specifically, lattice-aligned deposition gives exponential approach to jamming, whereas continuum, so-called "car parking" 1D deposition yields a power law, $\sim 1/t$, and these behaviors can actually be obtained by exact solution, e.g., [6]. More generally, convergence to jamming in 1D RSA can be understood by either directly studying the time-dependence of the coverage increase or be considering the behavior of the distribution of gaps [44] available for landing of the centers of particles the deposition attempts of which are not rejected (due to overlap with previously deposited particles). The "gaps" are generally regions of various possible shapes and orientations in more than 1D [45]. For discrete deposition, the "gap-size distribution" consists of delta function(s) representing available landing-site points, yielding exponential convergence for large times. For continuum deposition, the argument of Pomeau [44] has suggested that the gap size — measured by the length, $x$, into which a particle's center can land — distribution, $g(x)$, in 1D at large times is such that only the gaps that can fit a single particle dominate the dynamics. A further assumption of Pomeau [44] that $g(x)$ — defined such that $g(x)dx$ is the number of gaps of length between $x$ and $x + dx$ per unit length of the 1D line — is finite, non-singular at $x = 0$, yields the $1/t$ convergence.

Swendsen [45] extended this argument to more than 1D; see also [4,6] for arguments on how the continuum limit is obtained from discrete deposition of decreasing "mesh." In higher dimensions, particle shapes, rotational freedom vs. fixed orientation, and other "degrees of freedom" in particle positioning, as well as substrate patterning can all affect the approach to jamming. Specifically, this argument [45] for fixed-orientation hypercubes depositing on a continuous $d$-dimentional "substrate" suggests convergence to jamming according to $\sim (\ln t)^{d-1}/t$. This offers an interesting example of a deviation from a purely power-law behavior. When substrate is patterned — which has been a topic of recent interest, there is numerical evidence for both exponential and power-law convergence in 2D [10,11] for various geometries of the particles or surface pattern. However, in some cases the form of the convergence in 2D could not be numerically classified as power-law or exponential [10]. A recent study [9] of 1D models with "imprecise particle positioning" — represented by a pattern of lattice landing sites broadened into intervals — has offered analytical arguments for both fast and slow convergence depending on the model parameters. We note that, theoretical studies have also included extensions of the RSA model to allow particle motion, detachment, and other dynamical effects, as well as surface heterogeneity, disorder, and non-uniformity, e.g., [7,46-57].

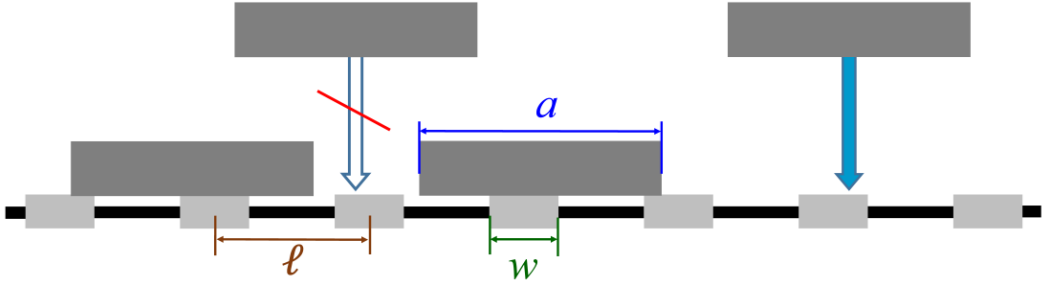

**Figure 1.** Model showing random sequential adsorption with particle centers only depositing in landing intervals of width $w$ centered at the sites of a 1D lattice of spacing $\ell$. Deposition of an incoming object aimed at a $w$-interval can be rejected (shown by the crossed arrow) because of overlap, here with two earlier-deposited objects. An allowed-deposition configuration is shown on the right.

In this work, we consider the model [9] illustrated in Figure 1. Here a uniform flux of particles of length $a$ reaches landing regions on a linear 1D substrate, but particles can only attach if they do not overlap one or two earlier-deposited particles. Substrate patterning is represented as follows: Particles can only be deposited if their centers fall within landing intervals that are broadened sites of a lattice of spacing $\ell$. These intervals have length $0 \leq w \leq \ell$, which extrapolates between lattice ($w = 0$) and continuum ($w = \ell$) deposition. This model was introduced in [9], were analytical arguments were presented for that, the convergence to jamming can be fast, exponential-type in some regions of the parameter space of varying $a/\ell$ vs. $w/\ell$, and it can be slow, power-law-type in other regions. Here we revisit a Pomeau-type argument and adapt it for the present model. We then report a numerical study of convergence properties as the model parameters are varied. There are regions in the parameter space of this model with the standard exponential, $\sim e^{-\text{const}\cdot t}$, or power-law, $\sim 1/t$, 1D convergences to jamming. However, our main finding in this work has been that, we also identified a segment with the $\sim 1/t^2$ convergence, and a region with the $\sim e^{-\text{const}\cdot t}/t$ convergence. We relate these time-dependent properties to new types of the gap-size distribution at the smallest reachable (for deposition) gap sizes, which offers a classification that extends the original Pomeau argument.

## 2. Definition of the Model and Extension of the Gap-Size Distribution Argument

In this section, we describe the model and use it as the framework to introduce analytical arguments pertaining to the properties of the gap-size distribution $g(x)$ and their implications for the asymptotic convergence to jamming. In the present model [9], it is convenient to consider a constant rate, $R$, of objects that attempt to deposit per each site of the 1D lattice, where these sites are broadened into intervals of length $w$ (Figure 1). Thus, we assume that the relevant particle flux, $\Phi$, is limited to particle center arrivals and deposition attempts only within the $w$-intervals and is given by

$$\Phi = R/w. \tag{1}$$

Arguments presented in [9] suggest that in the parameter space of $w/\ell$ and $a/\ell$, there is a periodicity of the *types of convergence* to jamming with period 1 in $a/\ell$. Most of our numerical studies, reported in Section 3, were for the region $1 \leq a/\ell \leq 2$, shown in Figure 2 (excluding the fully discrete deposition case, $w = 0$). It has been argued [9] that within triangle ABD the convergence to jamming should be fast (exponential type), whereas in the remaining sub-regions the convergence should be slow (power-law type). Special behaviors are found at certain lines and points, as addressed later. These properties, but of course not the actual values of the time-dependent coverage, are repeated for regions defined by

$$k - 1 < a/\ell \leq k, \tag{2}$$

where $k = 1, ...$ are integers; Figure 2 corresponds to $k = 2$.

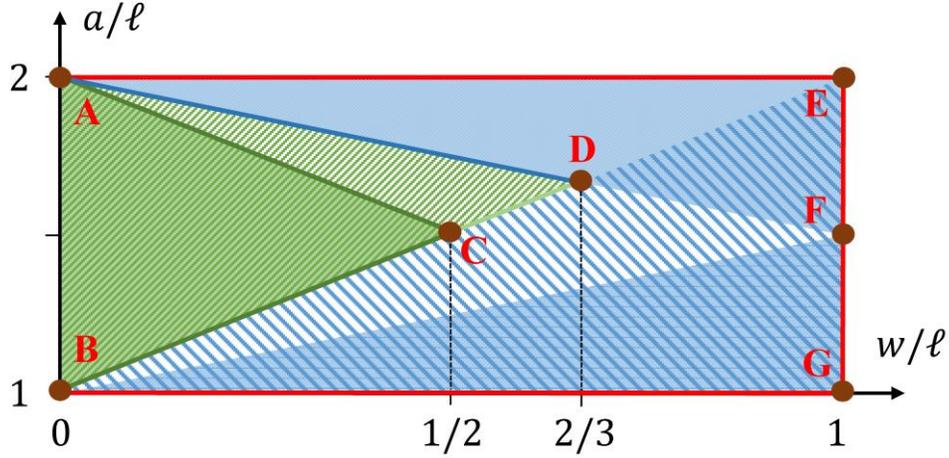

**Figure 2**. Different regions of convergence to jamming and certain other properties as described in the text. (Note that points C and F are at $a/\ell = 3/2$, whereas point D is at $a/\ell = 5/3$.)

The original Pomeau argument for a continuous-substrate deposition ($w = \ell$) involves an assumption that after a certain time, $t_0$, RSA dynamics can be considered dominated by deposition in gaps that can only fit a single object. For earlier times, $t \leq t_0$, there is a non-negligible number of large gaps that can fit more than a single object. Specifically, on deposition of an object in a gap, that gap can be either eliminated (if it is not overly large), or discontinuously shortened to a single or to two smaller gaps available for further deposition. Once only single-object-fitting gaps are considered, for $t > t_0$, the gap-size evolution involves only their elimination from the distribution, which is a *local* dynamics in the variable *x*, which denotes the length of the interval for landing of an object's center. Then the Pomeau argument for the emergence of the $1/t$ convergence, reviewed shortly, can be used.

In our model, the gap sizes for deposition of object centers are obviously limited to $x \leq w$. For simplicity, we will only consider the case $w \leq a$, because all our numerical results, see Section 3, were obtained in the regions of the parameter space with this condition satisfied. Then deposition of an object with its center in a gap completely blocks that particular gap for further deposition. However, for short times each deposited object can, depending on the specific values of *a*, partially or fully block one or two (for our range of *a* values) nearby *w*-gaps on both of its sides. Therefore, we still have non-local dynamics, including discontinuous shortening of some gaps, and also "locally correlated" events: blocking of more than a single *w*-gap by a single deposited object. Thus, in our model the Pomeau time $t_0$ is defined by that, the dynamics for $t > t_0$ is dominated by those gaps that are isolated, i.e., all their neighboring gaps were already blocked at earlier times by previously deposited objects. They will then undergo single-gap blocking events whenever an object's center is aimed at them during deposition.

Let us define the coverage, $\theta(t)$, as the average number of objects (or object centers) deposited per each lattice interval of length $\ell$. Note that the fraction of the covered length (of the 1D substrate) is then $a\theta(t)/\ell$. Both quantities are dimensionless, whereas the coverage as the density of objects per unit length is $\theta(t)/\ell$. Initially the substrate is assumed empty, and the coverage monotonically increases with time to its asymptotic infinite-time value

$$\theta_\infty = \theta(t = \infty) . \tag{3}$$

For times $t \geq t_0$, as the isolated gaps of size *x* are eliminated with the objects' arrival rate $\Phi x$, the gap-size distribution function, introduced in Section 1, changes according to

$$g(x, t \geq t_0) = g(x, t = t_0)e^{-\Phi x(t-t_0)} . \tag{4}$$

Following Pomeau [44], we now note that the total density (per unit length in 1D) of the objects that are deposited *after* a time $t \geq t_0$, is

$$[\theta_\infty - \theta(t \geq t_0)]/\ell = \int_0^w g(x, t \geq t_0)dx , \tag{5}$$

yielding

$$\theta(t \geq t_0) = \theta_\infty - \ell \int_0^w g(x, t_0) e^{-\Phi x(t-t_0)} dx \ . \tag{6}$$

Equations 4, 5, and 6 are of course approximate expressions that rely on the assumptions made for the properties of the large-time gap-size distribution. It is obvious from Equation 6 that, for large times, $t \gg t_0$, the integral that determines the convergence to the jamming coverage can be evaluated from the behavior of $g(x, t_0)$ near the smallest $x \geq 0$ value for which this function is non-zero.

Let us now consider various possible gap size distributions, starting with the case of the Pomeau argument [44] for continuum deposition. In our case, this corresponds to segment [EG]; see Figure 2. Here and below we use the standard bracket vs. parenthesis notation to respectively include or exclude the end points of straight-line segments. With the Pomeau assumption that the gap size distribution is finite at $x = 0$, $g(x, t_0)$ in Equation 6 can be replaced by a constant, $g(0, t_0)$, and the upper limit of integration set to $\infty$ in evaluating the convergence to jamming. Thus, we get the result

$$\Delta\theta \equiv \theta_\infty - \theta(t \gg t_0) \approx \frac{g(0,t_0)\ell}{\Phi t} = \frac{g(0,t_0)\ell w}{Rt} \ . \tag{7}$$

Arguments presented in [9] suggest that the behavior along the segments (AE] and (BG], see Figure 2, exclusive of the points A and B at $w = 0$, is exactly the same as for continuum deposition. Consideration, reported in [9], of how objects exclude each other, and how objects' center deposition is limited by the finiteness of the *w*-intervals suggest two partly overlapping (in the parameter space of *w* and *a*) mechanisms for very small gaps being possible in differently blue-color-coded regions in Figure 2: These regions jointly cover the interior of the shape AEGBD (a nonconvex pentagon). Indeed, our numerical study reported in Section 3, of gap-size distribution and also time-dependence of the coverage confirms that the Pomeau-type leading order behavior applies in this whole region. We note that, for asymptotic evaluation for times $t \gg t_0$, the gap-size distribution can be replaced by $g(0, t_0)e^{-\Phi xt}$, see Equation 3. This *x*-dependence is shown in Figure 3 for increasing time values.

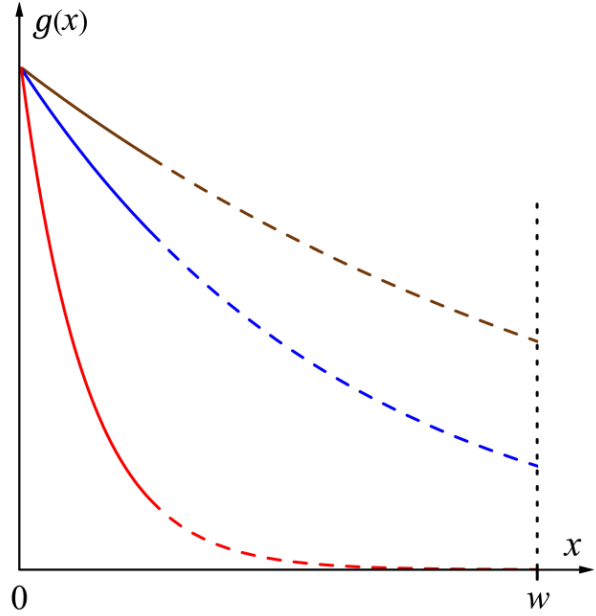

**Figure 3**. This figure schematically illustrates the function $g(0, t_0)e^{-\Phi t x}$, which can be used to approximate the gap-size distribution for small $x$ and large times, here $t_0 \ll t_{\text{brown}} < t_{\text{blue}} < t_{\text{red}}$, for evaluation of the Pomeau-type asymptotic convergence to jamming. We note that for larger $x$-values, marked schematically by the dashed-line extensions, the gap-size distribution can have significant variation as a function of $x$, which is not shown here (but see Section 3), including peaks. Furthermore, for our model a delta-function contribution is present at $x = w$, shown by a dotted vertical line. However, these features all decay exponentially and only contribute to transients in the time dependence of the coverage.

The other standard approach to jamming is exponential convergence. This case is best illustrated by $w = 0$, which corresponds to discrete deposition (at point-like landing sites for object centers). Obviously, point B (see Figure 2) corresponds to monomer deposition, whereas the segment [AB), including the end-point A, corresponds to dimer deposition. Arguments on how objects exclude each other [9] suggest that in fact, the whole triangle ABC, including its boundary except point B, also corresponds exactly to dimer deposition in the full time dependence of the coverage, when defined as the number of objects per lattice site, our $\theta(t)$. The gap-size distribution is then a delta-function, as shown schematically in Figure 4, with the amplitude decreasing exponentially with time for $t \gg t_0$: The integrand in Equation 6, $g(x, t_0)e^{-\Phi x(t-t_0)}$, can then be approximated by $\Theta \delta(x - w)e^{-\Phi wt}/\ell$, where $\Theta$ is a dimensionless coefficient. For the present region, ABC, we thus get

$$\Delta\theta(t \gg t_0) \approx \Theta e^{-\Phi wt} = \Theta e^{-Rt} . \tag{8}$$

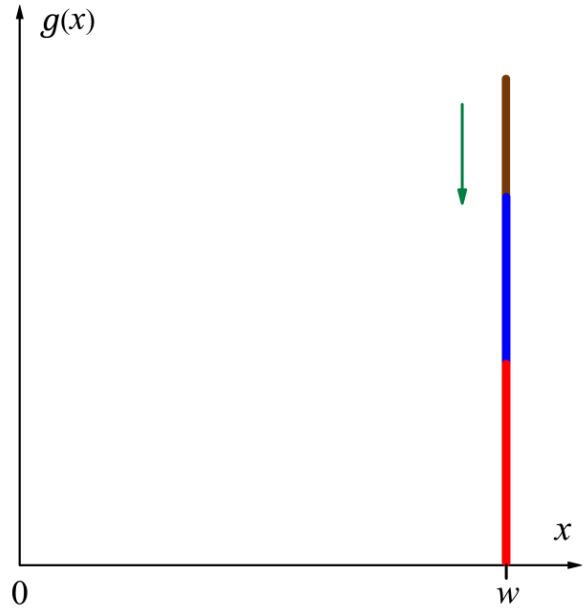

**Figure 4.** A schematic of the gap-size distribution function for discrete deposition. The distribution is a delta-function at $x = w$, which, when substituted in Equation 6, gives exponential convergence to jamming. Here $t_0 \ll t_{\text{brown}} < t_{\text{blue}} < t_{\text{red}}$.

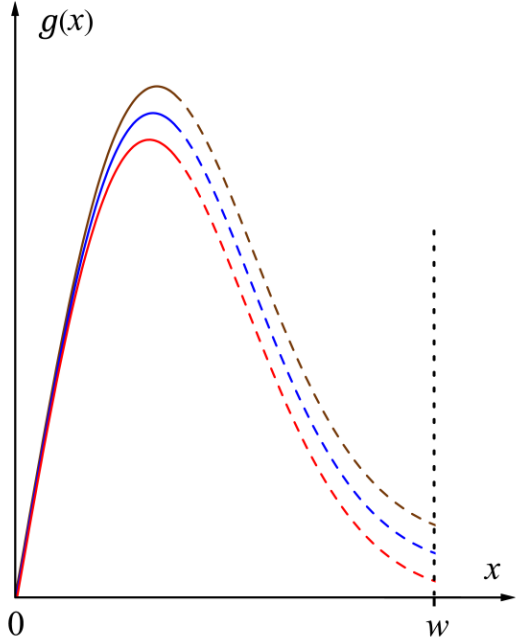

**Figure 5.** This figure schematically illustrates the function $\gamma x e^{-\Phi t x}$, which can be used to approximate the gap-size distribution for small $x$ and large times, here $t_0 \ll t_{\text{brown}} < t_{\text{blue}} < t_{\text{red}}$, for evaluation of the new type of asymptotic convergence to jamming found for segment (AD], see Figure 2. As in Figure 3, the behavior for larger $x$-values is marked schematically by the dashed-line extensions: The gap-size distribution then can have significant variation as a function of $x$, not shown here, and a delta-function contribution at $x = w$, the latter shown by a dotted vertical line. These larger-$x$ features all decay exponentially with time and only contribute to transients.

We now turn to *two new gap-size distribution types* suggested by our numerical simulations described in Section 3, which are non-standard and correspond to new asymptotic laws of approach to jamming in 1D. The first non-standard case is obtained for process parameters corresponding to segment (AD], including the end point D, but excluding the point A; see Figure 2. The same behavior is expected for such segments in regions bounded by other eger pairs of $a/\ell$ values, Equation 2, not shown in Figure 2, which replicate the considered region in the $(w/\ell, a/\ell)$ plane. In this case, it transpires that while the Pomeau assumption that the gap size distribution extends to $x = 0$ is applicable, the function $g(x,t)$ vanishes $x = 0$. Therefore, for asymptotic evaluation for times larger than $t_0$, the integrand in Equation 6 can be approximated by $\gamma x e^{-\Phi xt}$, see Figure 5, where $\gamma$ is the slope (the *x*-derivative) of $g(x, t = t_0)$ at the origin. The new asymptotic behavior, never before found for 1D RSA, is thus $\sim 1/t^2$,

$$\Delta\theta(t \gg t_0) \approx \frac{\gamma\ell}{(\Phi t)^2} = \frac{\gamma\ell w^2}{R^2 t^2}. \tag{9}$$

It is important to emphasize that the original Pomeau argument [44] involves several assumptions. Some of these apply as approximations starting from a certain large time $t_0$: The gap-size distribution's evolution in uncorrelated, i.e., only single gaps are eliminated in each deposition event. Furthermore, the evolution is local, i.e., no shorter gaps available for deposition are formed as byproducts of deposition events. These assumptions are used for evaluation of the asymptotic convergence to jamming, and they must be verified by numerical means or exact solutions. For the new behaviors proposed here, the same assumptions are still required. There is also the second conjecture of Pomeau [7], which assumes the finiteness at $x = 0$ of the function $g(x, t_0)$ at the "Pomeau starting time," $t_0$, which is set by the events for $t < t_0$ that eliminate larger gaps while shaping up the shorter-gap distribution: These deposition events' effects on $g(x, t)$ are non-local (can create shorter gaps) and correlated (can effect more than a single gap among those not yet fully blocked). Our "non-standard" behaviors therefore represent the deviation from the latter of the Pomeau assumptions: On the form at $t_0$ of the distribution for the smallest *x*-values that contribute to time-dependence of the type obtained from Equations 5, 6.

The second non-standard result that we found, corresponds to the points lying within of the triangle ACD, see Figure 2, including the segment (CD) without its end points, but excluding the other two boundaries, [AC] and [AD]. For this triangle, the gap-size distribution has a threshold property at small *x*, as shown in Figure 6, jumping from 0 for $x < h$ to a finite value at a certain threshold $x = h > 0$. This threshold can be easily determined from geometrical constrains on how depositing objects can partially block *w*-intervals during the early-stage deposition mentioned in the preceding paragraph; we get

$$h = 2k\ell - 2a - w, \tag{10}$$

for the interior of triangle ABC in the considered-region ($k = 2$) and its counterpart triangles in other regions defined in Equation 2. Similar to the Pomeau argument, here we found numerically that the *value* of $g(x = h, t_0)$ is non-zero, and thus the following expression can be used for asymptotic evaluation,

$$\Delta\theta(t \gg t_0) = \ell \int_h^w g(h, t_0) e^{-\Phi xt} dx. \tag{11}$$

The function $g(h, t_0) e^{-\Phi tx}$ is shown in Figure 6, and Equation 11 suggests the asymptotic convergence

$$\Delta\theta(t \gg t_0) \approx \frac{g(x=h,t_0)\ell w}{Rt} e^{-\Phi ht}. \tag{12}$$

This new asymptotic behavior involves a typical threshold-induced exponential, but modified by a power law. Note that the coefficient $g(x = h, t_0)$ and other such prefactors should actually be functions of the values of the problem parameters, such as $w$ and/or $a$, as further addressed in Section 3.3.

Interestingly, in the present model we did not find certain other forms of behavior of the gap-size distribution function that could be conjectured as possible. These include higher than linear-power vanishing at $x = 0$, or any (linear or higher-power) vanishing at the threshold $x = h$. For example, if the gap size distribution in some 1D model would have a threshold, i.e., vanish for $x < h$, but grow linearly, $\propto (x - h)$, for $x > h$, then we would have the asymptotic behavior $\Delta\theta \sim e^{-\Phi ht}/t^2$. We did not find any such convergences in our model, but we emphasize that our study, reported in Section 3, is entirely numerical, and therefore all conclusions are only as precise as allowed by the quality of the computer-generated data.

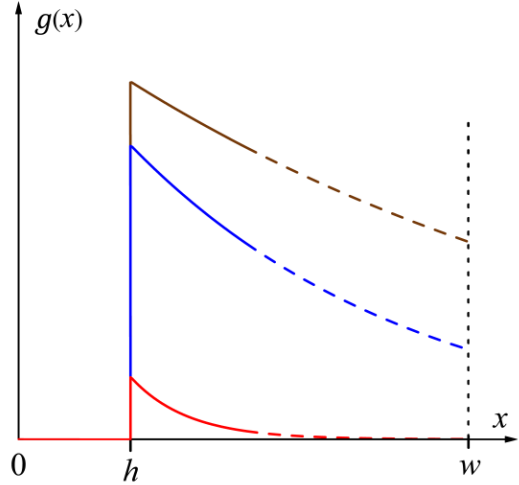

**Figure 6.** This figure schematically illustrates the function $g(h, t_0)e^{-\Phi tx}$, which can be used to approximate the gap-size distribution for small positive $x - h$ and large times, here $t_0 \ll t_{\text{brown}} < t_{\text{blue}} < t_{\text{red}}$, for evaluation of the new type of asymptotic convergence to jamming found for triangle ACD, see Figure 2. As in Figures 3 and 5, the behavior for larger $x$-values is marked schematically by the dashed-line extensions: The gap-size distribution then can have significant variation as a function of $x$, not shown here, including a delta-function contribution at $x = w$, the latter shown by a dotted vertical line. These features all decay exponentially with time and only contribute to transients. The gap-size distribution vanishes for $x < h$.

## 3. Numerical Simulations and Discussion of Results

### 3.1. General Comments on the Numerical Approach

Our numerical simulations were carried out by a direct Monte Carlo (MC) approach, without utilizing the acceleration techniques developed, e.g., in [4,58-60] for large time RSA and related models. The reason for this has been that acceleration techniques for RSA to a certain extent rely on system's configurations at large times reaching a Pomeau-type regime whereby all or most remaining gaps are covered in an uncorrelated and local fashion, as described in Section 2. We wanted to avoid any such assumptions, and therefore we used the most direct approach possible: randomly generated flux of particle centers aimed at *w*-intervals and deposited provided there is no overlap with already deposited particles. Our typical system sizes were up to $L = 500$, with periodic boundary conditions, where from now on we measure all lengths in units of $\ell$, and times in units of $1/R$, unless keeping the explicit $\ell$- or $R$-dependence is useful. It is expected that correlations in RSA are extremely short-ranged [6,61,62], and therefore moderate system sizes should suffice to avoid any noticeable finite-size effects [63]. However, we did notice non-negligible (as compared to statistical noise) finite-size effects at system sizes below $L \simeq 300$, which is likely attributable to that we have values of *a* not commensurate with $\ell$ and therefore with its multiple, *L*. We did not explore this effect in detail, because size effects do not affect the general properties of the large-*L* gap-size distribution.

All our simulation results were averaged over $10^4$ (some up to $10^6$) independent MC unit-time steps. Here each *unit MC time step* is defined to have on average 1 random deposition attempt per each *w*-interval. This means that our MC time is actually the dimensionless combination $Rt$. Considering our emphasis on the new time-dependences that represent approaches to jamming faster than the ordinary $\sim 1/t$ RSA power law, these times are sufficient for our purposes, as illustrated in the following subsections.

Besides measuring the time dependence, we also directly probed the gap-size distribution. Usually in RSA, the time dependence results are more definitive, because the gap-size distribution is much more sensitive to statistical noise. This is made apparent if we rewrite the expected, Pomeau-regime large-time relation, Equation 5, in the form

$$\Delta\theta(t \geq t_0)/\ell w = \int_0^w g(x, t \geq t_0)dx / w .  \qquad (13)$$

Then the right-hand-side expression indicates that the time-dependence of the convergence to jamming is simply the average of the function $g(x, t)$ over $0 \leq x \leq w$. We found it useful to use another type of averaging, defined later, which focuses on the small-$x$ behavior, but still averages the noise in $g(x)$.

We actually performed a rather large number of simulation for points within each region of Figure 2 identified for specific asymptotic behaviors in Section 2, and we also checked that similar asymptotic behaviors are numerically found for points in the smaller-$a$ region $1 \geq a > 0$, not shown in Figure 2, but discussed in [9]. The latter simulations are not detailed here; they were actually limited to the sub-region $1 > a \geq w$. Our main focus was on $2 \geq a > 1$ (Figure 2), because for $a > \ell$ ($\geq w$), deposition in a $w$-interval always blocks *at least* a part of one of the two nearest $w$-intervals. This choice is of relevance because in applications involving systems modeled by RSA, object sizes typically exceed, but not significantly, or are comparable to surface feature sizes. In the rest of this section we report representative numerical results for various regions of convergence identified in Section 2.

### 3.2. Quadratic in Time Power-Law Convergence to Jamming

In this subsection we consider the segment (AD], not including the $w = 0$ end-point A; see Figure 2. The jamming coverage along this segment was estimated numerically as described below. We note that arguments of the type presented in [9] establish that the jamming coverage in triangle ABC, inclusive its whole boundary except the point B, is the same as for lattice dimer deposition:

$$\theta_\infty^{\text{dimer}} = (1 - e^{-2})/2 \simeq 0.4323 . \tag{14}$$

As one changes the parameters to move into triangle ACD, the coverage gradually slightly decreases below the value in Equation 14, because within this triangle deposited objects can exclude parts of nearby landing intervals and therefore somewhat reduce the probability of dense configurations as compared to those for dimer deposition. On reaching the segment (AD], considered here, the jamming coverage is reduced to about

- 16 -

$$\theta_\infty^{(AD]} = 0.4301 \pm 0.0001 \,, \tag{15}$$

based on our MC data of the type illustrated in Figure 7. Furthermore, within the numerical accuracy and statistical noise, the whole time-dependence of the coverage along (AD] was found the same, independent of the values of $0 < w \leq 2/3$, with $a = 2 - (w/2)$, discontinuously jumping to dimer-deposition coverage at $w = 0$. These numerical observations suggest that there might exist an exact length-rescaling argument along segment (AD], similar to that described in [9] for segment (AE], that proves that the process is the same along the segment, but discontinuous at point A. We further discuss such numerical findings for the whole triangle ACD in the next subsection. However, we did not attempt to identify possible analytical arguments, because our focus has been on the new asymptotic properties of the gap-size distribution, described next.

Verification of the power-law convergence to jamming $\sim 1/t^2$ by directly simulating the time-dependence of the coverage is shown in Figure 7. The shown data for several points along the segment (AD] are plotted vs. $1/(Rt)^2$. For comparison, we also showed on the same plot the exact solution for continuum deposition for parameter values at point E (Figure 2), as well as numerical data for continuum deposition at a randomly selected point along segment (AE]. We note a clear confirmation of the $1/t^2$ convergence, when compared the continuum deposition (which corresponds to the $1/t$ convergence, here resulting in infinite slope at the origin). As mentioned earlier in connection with Equation 15, numerically the data for all the studied points on the segment (AD] are equal up to statistical noise, suggesting that the coverage is the same as a function of time. The inset in Figure 7 shows statistical noise in the data for points along (AD]. To further illustrate the statistical noise, numerical data for the point along (AE] was only averaged over $10^3$ runs and plotted with the exact solution curve (red color) behind it as the background. Interestingly, when the simulations were repeated with varying parameters $w$ and $a$ along (AD], with the same random-number sequence (obtained by the same initialization of the random number generator), then all the data sets collapsed to a single, but still noisy set. The equivalent of the inset would then be a single set of points with statistical noise. This offers further credence to the conjecture that the coverage is the same along (AD] for all times.

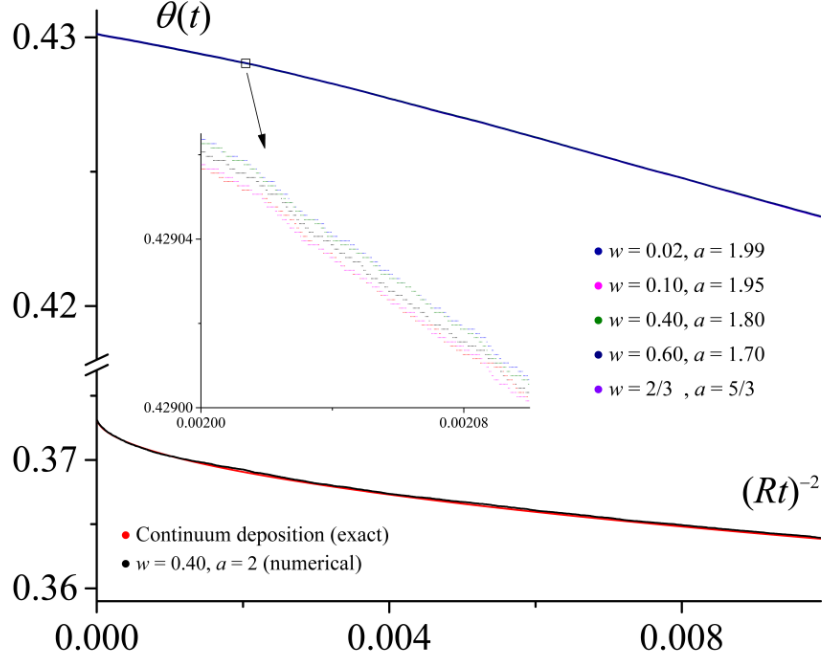

**Figure 7.** Illustrative numerical results for several points on segment (AD], plotted vs. $1/t^2$ to emphasize the new power-law behavior, as well as selected results for continuum deposition, shown for comparison. All the shown numerical data were obtained for lattice sizes $L = 500$, averaged over $10^4$ runs, except for the data for $w = 0.40$, $a = 2$ (see text for details and discussion). The inset illustrates the statistical noise in the data. Note that all the values in the legends (decimals or fractions) are exact.

Let us now consider the gap-size distribution, for which we show, in Figure 8(a), numerical data for three different MC times, for the parameter choice $(w,a) = (2/3, 5/3)$, i.e., for point D in Figure 2. This figure depicts the gap-size distribution averaged over small $x$-interval bins $\Delta x$. While statistical noise is clearly significant, these data estimate the function $g(x)$. The delta-function contribution to $g(x)$ at $x = w$, when averaged over an interval $\Delta x$, contributes a finite value $\sim 1/\Delta x$, and these values are shown as diamond symbols for the $x$-intervals' choice $\Delta x = 10^{-4}\ell$, which was used for Figure 8(a). Recall that some quantities here are shown with the assumption, mentioned earlier, that we use units $\ell = 1$; in dimensionless units $g(x)$ is measured per $1/\ell^2$. As seen in Figure 8(a), and this is a general observation for various parameter values, for times which are not large enough there are significant peak(s) in the gap-size distribution, which contribute(s) to transients in the time-dependence of the coverage, but decay exponentially for large times. The evolution of the distribution is then obviously non-local. For example, for time $t = 17$ (in units of $1/R$), one can see a clear buildup of the small-$x$ part of the distribution as compared to time $t = 12$. However, for time $t = 12$, we have a significantly larger delta-function contribution as compared to later times. Finally, for a large time, $t = 55$, there are too few gaps to see any statistically significant features.

As a further evidence of the new convergence law, we want to focus on the small-$x$ behavior of the distribution for very large times. However, with the available numerical capabilities there typically will then not be enough statistically significant data (too few gaps). Therefore, let us instead look at the time $t = 17$, see Figure 8(a), for which the distribution at small $x$ is already well formed, though a small number of larger gaps are still present. Even for this time, the actual data are noisy. Therefore, we use an averaging approach alluded to earlier, similar to Equation 13, but integrate only up to $x$, rather than up to $w$, i.e., we consider

$$\ell \int_0^x g(y) dy \approx \ell \gamma x^2 / 2 \:. \tag{16}$$

If $g(x)$ vanishes linearly at $x = 0$, then this integral should vanish quadratically, as shown in Equation 16, with the notation for the slope, $\gamma$, introduced in connection with discussion of Equation 9 and Figure 5. The square root of this integrated quantity is plotted in Figure 8(b) for time $t = 17$, and is strongly suggestive of that, indeed the expected behavior corresponding to linear vanishing of the gap-size distribution at the origin holds, which confirms the asymptotic $\sim 1/t^2$ convergence of the large-time coverage to jamming.

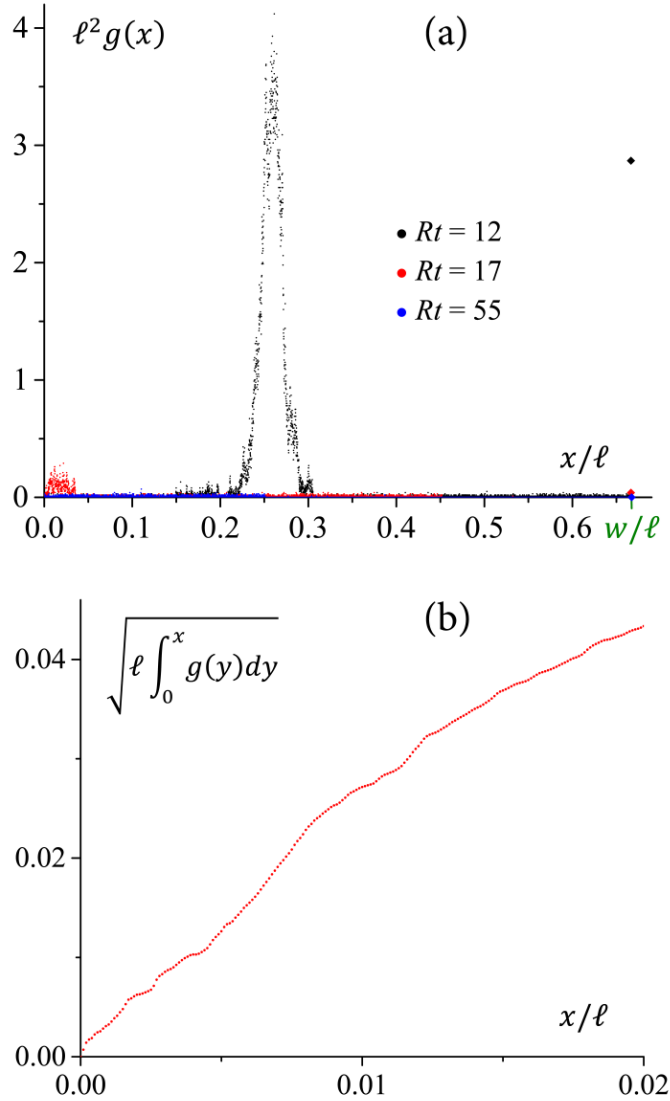

**Figure 8.** (a) Gap-size distribution for $(w,a) = (2/3, 5/3)$, obtained for $L = 500$, shown for three different MC times, as color-coded in the figure. The points marked by diamond symbols highlight the contribution of the delta-function at $x = w$. Note that the (black) symbol for $Rt = 12$ is at approximately at 2.87 vertically, whereas the larger-time (red and blue) diamonds are at much smaller $\ell^2 g$ values (0.04 and 0.00 respectively), which reflects the fast decay of the delta-function contribution. The buildup of the small-$x$ part of the distribution is seen for $Rt = 17$ (the red peak near the origin); see text for details. (b) The square root of the expression in Equation 16, numerically evaluated for time $Rt = 17$.

We also attempted to put error bounds on the power-law exponent in $\sim 1/t^p$, which is conjectured to have the value of $p = 2$ along (AD]. To this end, we obtained time-dependence data for two points along this segment, averaged over $10^6$ independent MC runs. Given the uncertainty in the value of $\theta_\infty^{(AD]}$ in Equation 15, by analysis of the log-log plots of the difference $\Delta\theta(t)$ vs. $t$, not detailed here, we could conclude that $p = 2.0 \pm 0.3$, which definitely excludes the value 1.

### 3.3. Convergence to Jamming for Gap-Size Distribution with a Threshold

In this subsection we consider the other new pattern of the behavior of gap-size distribution alluded to in Section 2: that with a threshold at $x = h$, with $w > h(a, w) > 0$ given by Equation 10 (here with $k = 2$). Interestingly, the threshold of the gap-size distribution is constant along segments parallel to AD within triangle ACD. However, the exponential time-dependence decay rate in terms of $Rt$, Equation 12, is set by the ratio $h/w$, which is actually constant along straight-line segments that connect point A with points on the side (CD) of the considered triangle ACD. Numerical simulations illustrated in Figure 9 not only confirm the expected behavior of the from $\sim e^{-(h/w)Rt}/Rt$, but also actually suggest that the time dependence along each of fixed-$(h/w)$ segments is the same for all the points on it, including the end-point on (CD), but excluding the end point A. This is similar to the behavior found along the segment (AD], with discontinuous jump in the coverage on approach to point A, and suggests the existence of a possible rescaling argument to establish this property exactly, not explored here.

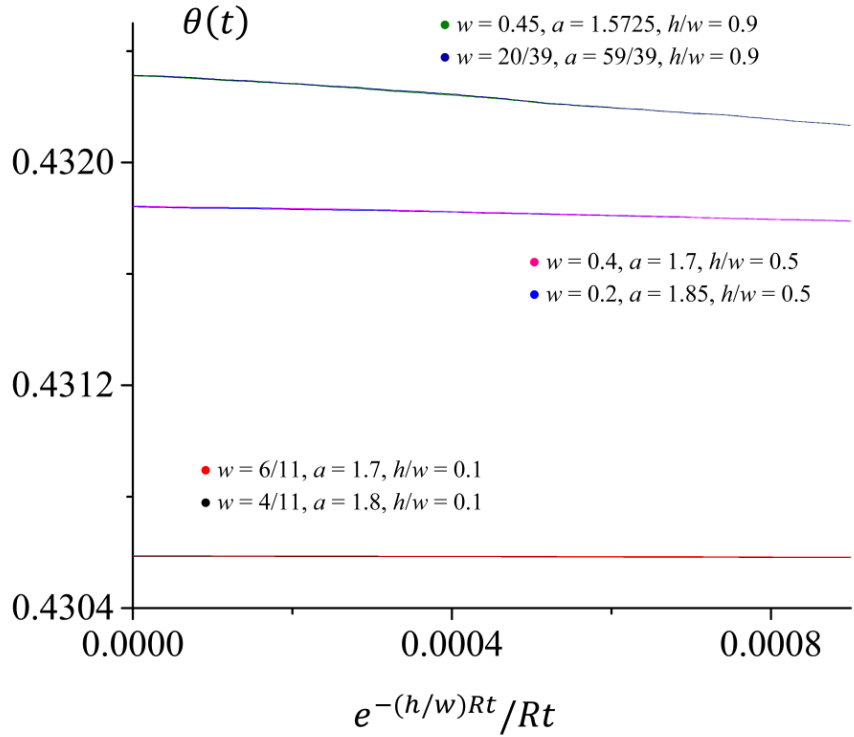

**Figure 9.** Data for time dependence of the coverage within triangle ACD, shown for pairs of points selected on three different segments that connect the point A with a point on (CD). There are total 6 data sets (dense points) shown, though they largely overlap pairwise. The top pair is for a segment close to AC, the bottom pair is for a segment close to AD, and the middle pair is for a segment in between. Note that all the values in the legends (decimals or fractions) are exact, and that the $w = 20/39$ point of the top pair of data sets was selected *on* segment (CD).

Note that the exponential form for the horizontal axis in Figure 9 places large times near that axis' origin, and constant-value behavior in the large-time limit confirms Equation 12. Since the values of the convergence rate, $h/w$, are different for the pairs of data sets in the figure, the actual displayed time intervals differ significantly. For example, the top data sets are depicted for times $Rt \geq 5.83...$, whereas the bottom data sets are shown for $Rt \geq 34.67...$, both corresponding to the same maximal value 0.0009 of the axis-variable. However, because of the exponential factors involved, the time 34.67... is at approximately $8 \times 10^{-16}$ along the horizontal axis for the top data sets. It is therefore not surprising that there are differences in the degree to which the constant behavior at the origin of this axis-variable is apparent for the three pairs of data sets. We also comment that the parametric $a$- and $w$-dependence of the prefactor $g(x = h, t_0; a, w)$ in Equation 12 should be such that the dimensionless product $\ell w g(x = h, t_0; a, w)$ has a constant value along the segments of fixed $h/w$, within triangle ACD.

The actual gap-size distributions for points within triangle ACD, including the segment (CD), were also studied and found rather noisy. However, there is a clear threshold at the expected value, $h = 4\ell - 2a - w$ (see Equation 10). To further explore the features of the distribution, we use the integrated quantity similar to that in Equation 16, but, considering the discussion at the end of the preceding paragraph, we replace the prefactor with $w$,

$$w \int_0^x g(y) dy \approx w g(x = h, t_0) \, e^{-(h/w)Rt} (x - h) \,, \tag{17}$$

where the right-hand side is for small $x - h \geq 0$. Figure 10 shows the left-hand side of Equation 17 calculated similarly to Figure 8(b). The data are points of gap counts within small $x$-axis bins, $10^{-4}\ell$, for a relatively large (considering the exponential convergence) time, $Rt = 17$. The data for $h = 0.1$, $w = 0.2$ show the threshold at the appropriate $h$-value as well as other expected features. Specifically, the curve saturates when the difference $x - h$ is no longer small. We have very few gaps in this regime; therefore the actually smooth saturation here looks made of small steps. The integration (summation of the gap counts) was taken past $x = w$, beyond which there are no additional gaps, but this allows us to see the discontinuous jump at $x = w$ due to the delta-function contribution. The main feature of interest besides the expected threshold property is clear evidence of the linear in $x - h$ growth that confirms the expected asymptotic properties discussed in connection with Equations 11 and 12 in Section 2. Figure 10 also shows the initial buildup of the integrated gap count for $h = 0.2$, with the appropriately adjusted values of $a$ and of $w = 0.4$, to have the same exponential convergence rate $h/w$. As discussed above, the use of the prefactor $w$ in Equation 17 then should guarantee the same slope of the linear in $x - h$ growth for both data sets (at the same time instance), which was indeed found to be the case, with the accuracy of approximately 6% considering the noise in the data.

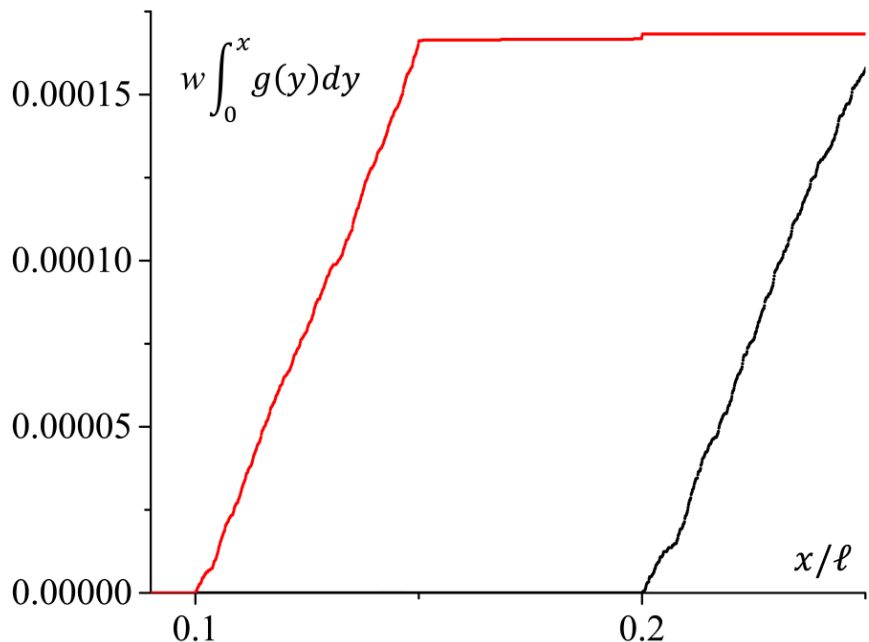

**Figure 10.** Integrated gap size distribution, Equation 17, for $(w,a,h) = (0.2,1.85,0.1)$: red data points, and for $(w,a,h) = (0.4,1.7,0.2)$: black data points, obtained for $L = 500$, shown for MC time $t = 17$, and averaged over $10^4$ MC runs. Note that the region of the top data set that shows saturation and looks made of small steps here due to very few gaps in it, starts not exactly at $x = 0.15$. As explained in the text, the step at $x = w$ ($= 0.2$) is real (it is due to the delta-function contribution). The second set of data has similar features, which are outside the shown range of $x$.

**3.4. Regions of Exponential and Standard Pomeau-Type Convergence, and Transients**

Let us now address the regions identified in Figure 2 that have standard RSA convergence patterns. Let us first mention triangle ABC, including its boundary. Our numerical simulations in this region confirmed the expected exponential convergence of the type shown in Equation 8, with the gap-size distribution as shown in Figure 4. In fact, this behavior can be derived exactly [9]: The time-dependence is that of the dimer deposition problem, except at point B, for which monomer deposition time-dependence applies. As mentioned earlier, we also studied numerically regions similar to those in Figure 2 for the the case of $k = 1$ (see Equation 2), limiting our study to the upper half of that region, for which $a > w$ (and avoiding the limit of point-like [64] objects, $a = 0$). All the considered types of behavior, though not the exact time-dependence, were found there as well.

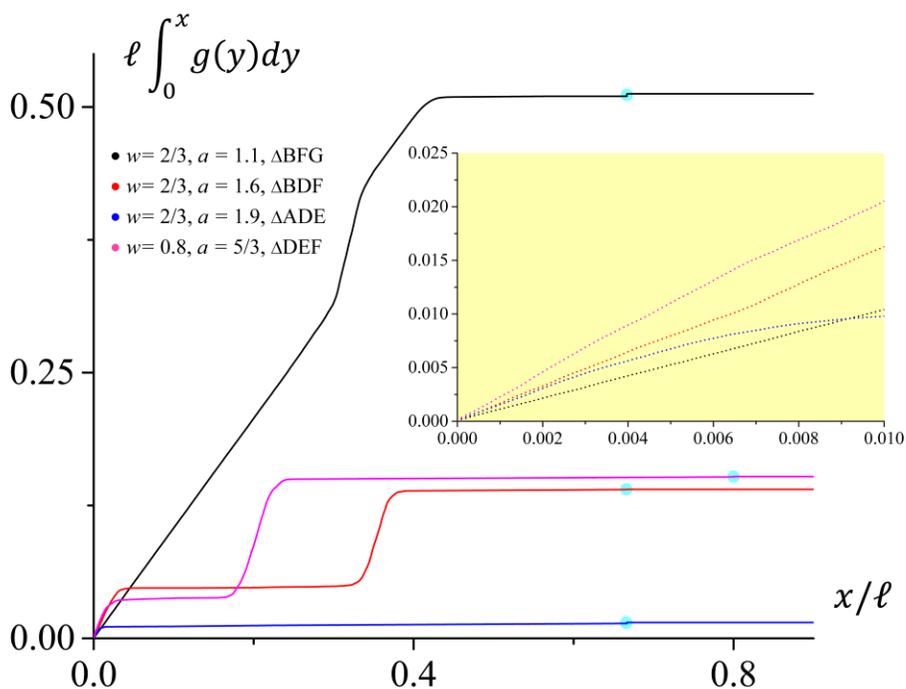

**Figure 11.** Integrated gap size distribution, Equation 18, for selected points within each of the triangular sub-regions of pentagon AEGBD, shown for MC time $t = 12$, averaged over $10^4$ MC runs, for system size 500. The Inset magnifies a small region near the origin, showing the expected linear behavior. The steps due to the delta function contributions at $x = w$ are highlighted by aqua-color circles.

The nonconvex pentagon AEGBD is of particular interest, because within it, as well as along the segments (AE], [EG], and (BG) we expect the standard Pomeau $1/t$ convergence.

Along the three afore-named segments the time-dependence of the fraction of the covered length, $a\theta(t)/\ell$, is exactly [9] that of the continuum car-parking problem. Furthermore, within the pentagon area there are sub-regions, quantified in [9] and differently blue-shaded/hatched in Figure 2, with one or two different mechanisms for the formation of arbitrarily small gaps. Such gaps can either be formed when a landing site for an object is "pinched" at both ends by two already-deposited objects, or when the "pinching" is by an already deposited object at one end, but by that the arriving object's center can only land very close to an end-point of a $w$-interval at the other end.

Generally, we found that the actual gap-size distributions at various points within the pentagon region are as noisy as the one shown in Figure 8(a), and they typically have a clearly identifiable but not necessarily symmetrical peak, centered at some $0 < x < w$, forming at intermediate times and then decreasing exponentially with time for large times, as well as the delta-function contribution at $x = w$. In some cases other features might be suggested by the data, but the data are too noisy for definitive conclusions. However, typically, clear evidence was found of that the gap-size distribution is *not* vanishing at $x = 0$. Figure 11 shows the integrated gap-size distribution defined as in Equation 16, which is here expected to behave linearly for small $x$ values,

$$\ell \int_0^x g(y)dy \approx \ell g(0, t_0) x \,. \tag{18}$$

Data are shown for selected points in each of the four triangular sub-regions of the pentagon; see Figure 2. We note that because of the slower, $1/t$, convergence, there are more gaps in this case, and, furthermore, we took a shorter time here than for Figures 8(b) and 10, which further ensures a larger gap count. As a result, in Figure 11 we can demonstrate the general features of the integrated gap-size distributions, including linear behavior at the origin (shown in the Inset), sharp-rise regions (corresponding to peaks), and steps at the delta function location at $x = w$, past which there are no more gaps and the integral is constant.

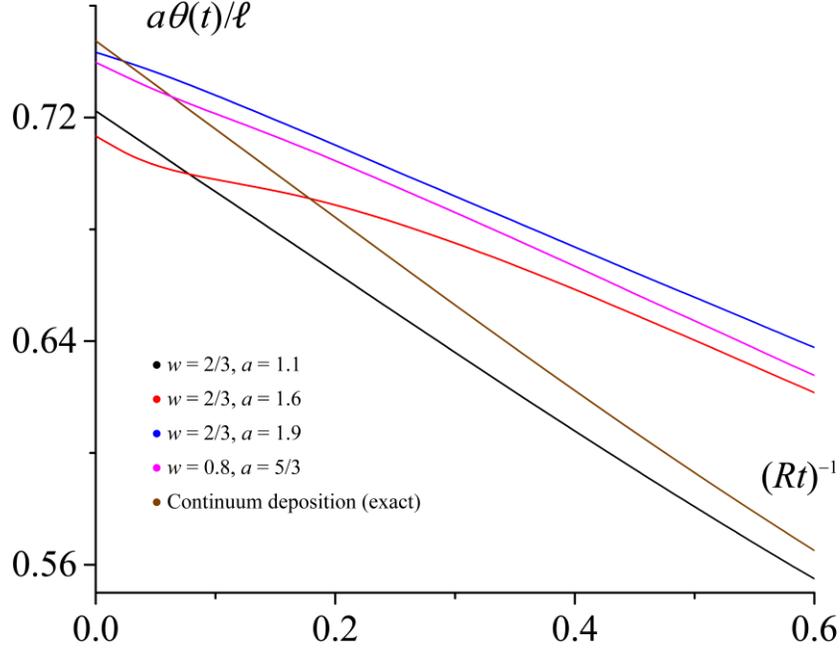

**Figure 12.** Time dependence of the coverage plotted as a function of $1/t$ for the same four points as in Figure 11, with the same color-coding. These data were obtained as in Figure 11, for times up to $10^4$. The data are shown for the fraction of the covered length, $a\theta(t)/\ell$, which facilitates comparison with the exactly solvable continuum-deposition model, the solution for which is also shown. This exact result applies along segments (AE], [EG], and (B,G] in Figure 2.

The time dependence of the coverage for the four parameter choices described above is shown in Figure 12. Generally, we see clear confirmation of that the $\sim 1/t$ behavior applies. For comparison, an exactly solvable model result [6] for continuum deposition is also shown. The plots in Figures 11, 12 illustrate that, as the parameters $w$ and $a$ are varied within the nonconvex pentagon AEGBD, we obtain different models with different time dependences and gap-size distributions. The only unifying feature is the standard Pomeau-type convergence. The values of the jamming coverage are clearly different, and they were estimated as $(a/\ell)\theta_\infty^{w=2/3,a=1.1} = 0.7224$, $(a/\ell)\theta_\infty^{w=2/3,a=1.6} = 0.7134$, $(a/\ell)\theta_\infty^{w=2/3,a=1.9} = 0.7473$, $(a/\ell)\theta_\infty^{w=0.8,a=5/3} = 0.7397$, all with the error bars of $\pm 0.0002$. This should be compared to the exact solution value [6] for continuum deposition, $(a/\ell)\theta_\infty^{\text{continuum}} \simeq 0.7476$. As in Section 3.2, we attempted to put error bounds on the power-law convergence exponent in $\sim 1/t^p$. Given the uncertainty in the values of $\theta_\infty$, analysis of the log-log plots of the difference $\Delta\theta(t)$ vs. $t$, not detailed here, suggests the estimate $p = 1.0 \pm 0.2$.

We did not find any systematic evidence for "crossover scaling" or other quantifiable mechanisms of how the convergence to jamming changes as the parameters $w$ and $a$ are varied to approach special regions associated with triangle ACD. It seems that the transients observed in the data are associated with various features of the gap-size distribution, notably, peaks, that build up at intermediate times, but then decay exponentially fast at larger times. However, no systematic trends were found within the scope of our present study. One good example is the point $w = 2/3, a = 1.6$; see Figures 12. It shows a more pronounced transient behavior than the other time-dependences shown. One could speculate that this is due to this point being closer to the special point D (Figure 2), but no systematic enhancement of this transient behavior was noticed for points even closer to D (within the pentagon AEGBD). Furthermore, we note that this point is within one of the two regions of the pentagon (out of the total four triangular regions within it, see Figure 2) that only have a single, rather than two different mechanisms for the formation of infinitesimally small gaps [9]. Therefore, the transient behavior cannot be attributed to two contributions to the gap-size distribution at small $x$.

## 4. Summary

Our main result has been the finding, based on numerical evidence, that part of the standard arguments [44,45] used to derive the convergence laws to jamming in RSA requires generalization. This applies specifically to the assumption that the large-time gap-size distribution is constant in the limit of very small gaps. We found examples of linearly vanishing (at small gap sizes) and threshold-type distributions. These findings suggest new asymptotic convergence-to-jamming laws, exemplified by those identified for our 1D model: Equations 9, 12. We emphasize, however, that our findings are entirely numerical, and additional studies are called for. Analytical progress (exact results) might be possible, as suggested by our numerical evidence that the time dependence of the coverage is the same along certain segments in the parameter space of the considered 1D model.